# Laboratory Experiments on Agglomeration of Particles in a Granular Stream

Yuuya Nagaashi[1]
Corresponding author
Email: yuuya.nagaashi@stu.kobe-u.ac.jp

Tomomi Omura[1]
Email: tomura@stu.kobe-u.ac.jp

Masato Kiuchi[1]
Email: mkiuchi06@gmail.com

Akiko M. Nakamura[1]
Email: amnakamu@kobe-u.ac.jp

Koji Wada[2]
Email: wada@perc.it-chiba.ac.jp

Sunao Hasegawa[3]
Email: hasehase@isas.jaxa.jp

[1] Graduate School of Science, Kobe University, 1-1 Rokkodaicho, Nada-ku, Kobe 657-8501, Japan

[2] Planetary Exploration Research Center, Chiba Institute of Technology, 2-17-1 Tsudanuma, Narashino, Chiba 275-0016, Japan

[3] Institute of Space and Astronautical Science, Japan Aerospace Exploration Agency, 3-1-1 Yoshinodai, Chuo-ku, Sagamihara, Kanagawa 252-5210, Japan

## Abstract

Inelastic collisions occur among regolith particles, such as those in the ejecta curtain from a crater, and may cause clustering or agglomeration of particles and thus produce discrete patterns of ejecta deposits around a crater. Previous studies have shown that clusters, and even agglomerates, are formed via mutual, inelastic collisions of spherical particles due to adhering forces between particles in granular streams. To investigate the condition of agglomerate formation in granular streams, we conducted laboratory experiments of granular streams using both spherical and irregular, non-spherical particles. Measurements of particle adhesion in this study were performed using a centrifugal separation method, in contrast to the previous study in which atomic force microscopy (AFM) was used. This enabled simultaneous measurements of multiple particles of various shapes for a statistical analysis of the results. With similar relative velocities and adhesion values, irregular particles were found to form agglomerates much more easily than spherical particles. The axial ratio of the agglomerates of spherical particles and irregular particles was similar and was in accordance with those observed in previous laboratory studies, whereas the size of the agglomerates of irregular particles was larger than the size of spherical particles. The degree of agglomeration and the size of agglomerates can be used as an indicator of the shape or adhesive force of the particles in granular stream. Our findings on agglomeration in granular streams could provide new insights into the origin of rays on airless bodies and grooves on Phobos.

## Introduction

The flow of granular material enables the formation of discrete structures due to inelastic collisions between particles. For example, high-speed camera observations of the ejection of glass beads and silica sand from craters recorded in laboratory impact experiments have revealed mesh-like patterns in the ejecta curtain (Kadono et al., 2015). Such patterns have also been reproduced using numerical simulations of particles with non-unity restitution coefficients (Kadono et al., 2015). It is possible that this process is relevant to the origin of rays around fresh craters on the Moon or other planetary bodies (Kadono et al., 2015).

The simplest granular flow is a stream of particles. A laboratory observation of free-falling streams of spherical particles has shown that uniform streams initially form clusters as they fall under Earth's gravity (Royer et al., 2009). The clusters are connected by bridges, and the bridges become thinner as the clusters fall downward, breaking apart from one another to form discrete clusters. A numerical simulation (Waitukaitis et al., 2011) has shown that particles forming droplets have a very small relative velocity and that the average coordination number (i.e., the number of contacts of each particle with other particles) is approximately 4. A previous laboratory experiment (Royer et al., 2009) showed that discrete clusters do not change their shape significantly after formation until they are destroyed via air drag; based on the above information, these are most likely agglomerates, not just clusters. Here, "cluster" is defined as a group of particles regardless of whether or not they adhere, and "agglomerate" is defined as a group of particles that adhere to one another. Henceforth, this paper refers to the groups of particles formed in granular streams as agglomerates, although Royer et al. (2009) refers to them as clusters. Royer et al. (2009) indicated that the formation of agglomerates is due to the adhesive force produced by a combination of van der Waals interactions and capillary forces between nanoscale surface asperities and not due to hydrodynamic interactions with ambient gas. It has also been experimentally implied that the formation of agglomerates is insensitive to the restitution coefficient of particles (Royer et al., 2009). A subsequent numerical study showed that a lower restitution coefficient of particles acts to increase collimation of the stream and that the adhesive force drives agglomerate formation (Waitukaitis et al., 2011), which agrees with laboratory experiments (Royer et al., 2009).

To obtain a better understanding of the conditions required for the formation of agglomerates in a granular stream and to apply the conditions of impact phenomena on asteroids and other celestial bodies, we conducted laboratory experiments of granular streams with an emphasis on irregular, non-spherical particles, which are a more suitable analogue of regolith particles than spherical particles are. As mentioned above, mesh-like patterns have been observed in the ejecta curtain in laboratory impact experiments (Kadono et al., 2015). Keeping this in mind, we turned our attention to the ejecta curtain produced by an artificial impactor, such as the small carry-on impactor (SCI) of the Hayabusa2 asteroid mission of ISAS/JAXA (Arakawa et al., 2017), or that was produced by natural collisions of meteoroids on the regolith surface of small bodies. The ejecta curtain from the surface of a small body is expected to show discrete structures, such as mesh patterns, clusters, or agglomerates of regolith particles. In this study, we attempted to demonstrate how the study of these structures could reveal valuable information about the physical properties of regolith particles.

## Experiments
### Particles
We used poly-dispersed spherical glass beads with median diameters of 55 μm (hereafter called 50-μm beads) and 93 μm (100-μm beads) (Fujikihan Co., Ltd.). To examine the effect of the shape on the agglomeration of particles in granular streams, we also used irregular alumina particles with a median diameter of 77 μm (The Association of Powder Process Industry and Engineering, JAPAN) and silica sand grains (Miyazaki Chemical Co., Ltd.) with a median diameter of 73 μm, which were prepared using a sieve with a 45-μm mesh opening to reduce fine particles, as in a previous study (Omura et al., 2016). Among the particles, the silica sand grains are the most representative of the regolith particles on asteroids in terms of composition, size, and shape (e.g., Omura and Nakamura, 2017). We also show the size distributions of the samples determined by two methods: using a laser diffractometer (SHIMAZU SALD-3000S) installed at Kobe University, and using an optical microscope (Nikon ECLIPSE ME600) and a digital camera (Nikon D300), shown in Fig. 1. Both size distributions are consistent. Fig. 2 shows scanning electron microscopy (SEM) images of the particles. The roughnesses of the glass bead were examined using a laser confocal refractometer. The vertical height of the surface was measured with 1 μm spacing and a vertical resolution of 0.1 μm (0.2 μm reproducibility is guaranteed by the supplier). According to Gadelmawla et al. (2002), the average roughness of a surface, $R_a$, is defined as follows:

$$R_a = \frac{1}{n}\sum_{i=1}^{n}|y_i| \quad (1)$$

where $y_i$ is the vertical deviation from the mean line. We approximated the mean curved line by a polynomial function. $R_a$ was 1.13 ± 0.24 μm for the 50-μm beads and 3.82 ± 0.45 μm for the 100-μm beads. The restitution coefficient of a glass sphere (diameter: 3.2 mm) against a 10 mm-thick glass plate measured using an acoustic emission sensor (Higa et al., 1996) was 0.957 ± 0.012 for a collision velocity in the range of 0.5–1.3 ms$^{-1}$. The restitution coefficient of an alumina sphere with a diameter of 4 mm against a 10-mm-thick alumina plate was also measured and found to be 0.975 ± 0.006. Table 1 summarizes the physical characteristics of the particles.

**Measurement of adhesive force**

In the previous study of granular streams (Royer et al., 2009), the adhesive force between micron-sized spherical silica particles was measured using an atomic force microscope cantilever. In this study, however, we used a centrifugal separation method (centrifugal method) (e.g., Krupp, 1967) to measure the adhesive force between a particle and a plate. This method can directly measure the adhesive force of multiple particles simultaneously, including those of different shapes, enabling a statistical analysis of the results.

All measurements of particle adhesive force were conducted in air. A schematic diagram of the experimental configuration used for the force measurements is shown in Fig. 3a, and the apparatuses are shown in Figs. 3b and 3c. First, we deposited particles over Petri dishes, or in some cases optical glass plates, using a cotton bud. Before particle deposition, the Petri dish surface was washed with ethanol. We then took an image of the particles adhering to the Petri dish using an optical microscope (Nikon ECLIPSE ME600) and a digital camera (Nikon D300). We were able to see that oil and other factors were not attached to the Petri dish. With this measurement approach it was not necessary to sieve particles because the size of each particle could be determined from the image. ImageJ software (Image Processing and Analysis in Java: https://imagej.nih.gov/ij/) was used to analyze the images. Fig. 4a shows an example of an image of a Petri dish with particles. Some particles were in contact with one or more particles; these particles were not included in the measurement results.

Next, the Petri dish was placed in a cylindrical basket with its surface normally aligned with the axis of the basket, as shown in Fig. 3a. The basket was then put in a table-top centrifuge (Kubota KN-70) at Kobe University, as shown in Fig. 3b. The distance from the axis of rotation to the surface of the Petri dish was set at 0.08 m or 0.12 m, according to the height of the gypsum filler shown in Fig. 3c. The basket was rotated at 300 rpm to apply a centrifugal force to particles adhered to the Petri dish. After a few minutes the rotation was stopped and the Petri dish was gently removed from the basket. We took an optical microscope image at the same location of the dish as before applying the centrifugal force, as shown in Fig. 4b. The Petri dish was then once again placed in the basket and rotated faster than before. This process was repeated, as the rotational speed was increased in 15 steps to 5,000 rpm, which was the maximum rotational speed of the centrifuge used in this experiment. The centrifugal force acting on the particles increased with the rotation speed. The particles subsequently separated from the glass plate when the resultant force of two forces, i.e., the centrifugal force and the force of gravity, which was approximately equal to the centrifugal force, became larger than the adhesive force between the particle and the dish. Measurements were conducted four times for each particle type. The number of detached particles for which the measurement was completed was 29, 98, 262, and 596 for the 50-μm beads, 100-μm beads, alumina particles, and silica sand grains, respectively. In the case of small particles, the centrifugal force applied was smaller than the adhesive force between the particle and the Petri dish; thus, 71, 0, 92, and 29 particles remained on the Petri dish for the 50-μm beads, 100-μm beads, alumina particles, and silica sand grains, respectively.

We measured the vibration acceleration of the centrifuge using an accelerometer placed on the top cover of the centrifuge and found that it was up to 0.21 G, which is considerably smaller than the centrifugal acceleration. In addition, because the basket could tilt freely from a vertical to horizontal position, the vertical direction of the Petri dishes should have been in the direction of the resultant force of the centrifugal force and gravity, which was approximately in the direction of the centrifugal force. Adhesive forces measured in this experiment included van der Waals, electrostatic forces, and the effects of moisture adsorption from the atmosphere on the particle surface. Regarding the electrostatic force, previous studies, such as that of Royer et al. (2009), involving granular streams indicated that this force is very small; as such, the electrostatic force contribution was assumed to be negligible in this study. As another concern, as indicated in the electron microscopy image of Fig. 2, it is conceivable that fine objects of a sub-micron to micron scale adhered to the particles; however, the essential point is that the adhesion of particles in this state was measured and that granular stream experiments in the air were also conducted with the particles in the same state.

The possible range of rotation speeds at which the particle separated from the plate was determined based on optical microscope images to estimate the adhesive force between the particle and the Petri dish. The mass of each particle, which was used for the calculation of the centrifugal force, was estimated using the size of the particle from the images and the mass density value. The optical microscope images were then translated into binary data using ImageJ software to measure the area based on the major and minor axes of each particle. For a glass bead, we assumed the shape to be a perfect sphere and determined the radius from the projected area, which enabled calculation of the volume. For an irregularly shaped particle of either alumina particles or silica sand, we assumed that the shape was instead ellipsoidal. The axial ratios of the two-dimensional (2D) projections were 0.63 ± 0.16 with a median of 0.64 and 0.66 ± 0.16 with a median of 0.67 for the alumina particles and silica sand grains, respectively. These values are smaller than the value of the medium to the maximum axial ratio (0.73) and larger than the minimum to the maximum axial ratio (0.50) of collisional fragments (Fujiwara, 1978). We assumed that the axial ratio of the irregularly shaped particles was similar to that of collisional fragments as $2c : \sqrt{2}c : c$, where $c$ is the minimum axis. We determined the value of $c$ from the projected area $S$ of the microscope image and calculated the volume of particle $V$ on the assumption that the minimum axis was along the line of sight, i.e., that the minimum axis was vertical to the Petri dish:

$$S = \frac{\pi c^2}{\sqrt{2}}, (2)$$

$$V = \frac{\sqrt{2}\pi c^3}{3}. (3)$$

Using this method, we were able to estimate the centrifugal force $F_{cen}(\omega)$ applied to each particle using the calculated volumes together with the density of the material (shown in Table 1), the distance from the axis of rotation, and the rotation speed $\omega$. The adhesive force between the particle and the Petri dish was then obtained as the geometric mean of the two subsequent centrifugal forces applied to the particle as $\sqrt{F_{cen}(\omega_i)F_{cen}(\omega_{i+1})}$, where $F_{cen}(\omega_i)$ denotes the force at the maximum rotation speed $\omega_i$ at which the particle remained on the Petri dish and $F_{cen}(\omega_{i+1})$ denotes the force at the rotation speed $\omega_{i+1}$ at which the particle no longer remained on the dish. Note that if we assume the extreme case of the maximum axis of a particle being along the line of sight when the optical microscope image is acquired, the volume estimate becomes $2\sqrt{2}$ times larger than the above. Therefore, adhesive forces may be a few times larger than the current estimates.

The surface roughness ($R_a$) of the Petri dish was 0.18 μm. We used two optical glass plates with surface roughnesses of 0.032 and 0.046 μm, respectively, to examine the effect of the surface roughness on the cohesive force and particle shape.

**Granular stream experiments**

The granular stream experimental setup is shown in Figs. 5a and 5b. The experiments were performed in a vacuum chamber with a diameter of 1.5 m and a height of 2.0 m in ISAS/JAXA and in a chamber with a diameter of 0.15 m and a height of 1.5 m at Kobe University. We changed the ambient pressure in the vacuum chamber to examine the effect on agglomeration. The ambient pressure was set to between 3 and $10^5$ Pa for the ISAS experiments, and between 10 and $10^5$ Pa for Kobe University experiments. A plastic funnel containing the sample particles was placed at a height of 1.8 m and 1.5 m from the bottom of the chambers for the ISAS and Kobe University experiments, respectively. The particles exited the bottom of the funnel through an aperture (diameter: 8 mm). Following the opening of an electromagnetically sealed shutter at the bottom of the funnel, the particles fell freely from the aperture as a granular stream.

We used a digital camera (Casio EXILIM PRO EX-F1) to take images (432 × 192 pixels) of granular streams in the ISAS chamber at 600 fps in air and at a reduced ambient pressure. An additional 6400-fps high-speed camera (Photron FASTCAM Mini AX) was used to acquire more detailed images (1024 × 1024 pixels) in air for the ISAS experiments. The cameras were initially held at the height of the aperture with an electromagnet. After the shutter of the funnel aperture was opened, the electromagnet for the camera was deactivated, and the camera was allowed to slide along two perpendicular shafts together with the stream. The granular stream was illuminated over its entire length using taped light-emitting diodes (LEDs). We used a high-speed camera (Photron FASTCAM SA1.1) fixed at vertical distances of 0 m and 0.9 m from the aperture to obtain high-resolution images (1024 × 1024 pixels) in the experiments conducted at Kobe University. The camera did not move along the granular stream. The spatial resolution was 0.02 mm/pixel, and the frame rate was 5400 fps.

## Results and Discussion
### Adhesive force

In Fig. 6, the ratio of the experimentally measured value of the adhesive force to the calculated theoretical reference value, described below, is shown as a function of the cumulative percentage of particles for which we determined the adhesive force. The range of the measured adhesive force is shown in Table 2. As mentioned previously, the centrifugal force generated at 5000 rpm in this study was not sufficient to separate all of the particles from the Petri dish for the 50-μm glass beads, while all of the 100-μm glass beads were separated successfully from the Petri dish before the maximum centrifugal force was achieved. The percentage of the residual 50-μm glass beads on the Petri dish was 71%; hence, the curve in Fig. 6a for 50-μm glass beads shows the lower limit of the adhesive force distribution. The percentage of the residual alumina particles and silica sand grains were 26% and 5%, respectively. Most of the residual particles and grains had small diameters of approximately 30 μm; thus, the curves for the alumina particles and silica sand in Fig. 6a can be regarded as representative for alumina particles and silica sand (approximately 75-μm particles).

We refer to the theoretical value of the adhesive force based on Johnson–Kendall–Roberts (JKR) theory (Johnson et al., 1971), although the capillary force is significant at atmospheric pressure, as discussed later. The force required to separate two spheres in contact, $F_c$, is given by

$$F_c = 3\pi\gamma R, \quad (4)$$

where $\gamma$ is the surface energy of the particle, and $R$ is the reduced radius of two spheres with radii of $R_1$ and $R_2$, respectively:

$$1/R = 1/R_1 + 1/R_2. \quad (5)$$

In this case, the contact is between a glass plate and a particle. Thus, $R$ is simply equal to the radius of the particle. We calculated the force for perfect spheres of radius $R$ for glass beads, alumina, and silica sand grains according to Eq. (4) as reference values. In the calculation, we used the radius $R$ for each particle determined from the 2D projected area by

$$S = \pi R^2. \quad (6)$$

We assumed a value of $\gamma = 0.025$ J/m² of $SiO_2$ particles (Heim et al., 1999) for the glass beads, silica sand, and the glass plate; $\gamma = 0.041$ J/m² was used for alumina particles (Burnham et al., 1990). We adopted the formula for the adhesive force, $\gamma = \sqrt{\gamma_1\gamma_2}$, between different materials with surface energies of $\gamma_1$ and $\gamma_2$ (Burnham et al., 1990). The adhesive forces between the Petri dish and the 50-μm glass beads, 100-μm glass beads, 77-μm alumina spheres, and 73-μm silica spheres were 5.9, 12, 12, and 8.6 μN, respectively.

Fig. 6a shows that approximately 80% of the 50-μm glass beads that detached from the Petri dish had an adhesive force that exceeded 10% of the theoretical reference value. If we take the percentage of residual glass beads on the Petri dish, the distribution of adhesive forces of the 50-μm glass beads had a higher median value than that shown in Fig. 6a (shown in Fig. 6b). On the other hand, more than half of the 100-μm glass beads had values less than 5% of the theoretical reference values of perfect spheres. From the different shapes of the curves of irregular particles compared with those of the glass beads in Fig. 6, it can be said that the adhesive force is likely influenced by the macroscopic shape of the particle. On the other hand, the data for the 100-μm glass beads were much smaller than the theoretical reference values compared with those for the 50-μm glass beads; this is likely due to the high roughness ($R_a$) of the 100-μm glass beads in comparison with the 50-μm beads. The compression length of a sphere (delta in JKR theory) is 0.54 nm for the 50-μm glass beads and 0.69 nm for the 100-μm glass beads. Compared with the one order of magnitude larger surface roughness, which is approximately 1 μm for the 50-μm beads and 4 μm for the 100-μm beads, it can be said that perfect contact was not achieved. Fig. 7 shows that the spread in the measurement values of the 100-µm glass beads with the optical glass plates was smaller than the case in which the Petri dish was used, implying that the microscopic surface roughness has a significant effect on adhesive forces.

We calculated the effective surface energy of particles, $\gamma_{\text{eff}}$, which reproduces the measurement value. Table 3 summarizes the range of the $\gamma_{\text{eff}}$ values of the particles corresponding to 10–90% of the total number of particles; note that the percentage of the residual particles on the Petri dish is not taken into account. The measured values in this study were considerably smaller than the reference values, based on the previous measurements of surface energy (Heim et al., 1999; Burnham et al., 1990). As mentioned above, this is probably because the contact between the particles and the Petri dish was less than perfect due to the high surface roughness. The root mean square deviation from the ideal sphere of a micron-sized $SiO_2$ particle used in the previous study was 0.13 nm (Heim et al., 1999).

**Horizontal particle velocity just below the aperture**

We analyzed the motion of the particles just below the aperture via particle image velocimetry (PIV), using the fixed high-speed camera images shown in Fig. 8. The analysis was only applied to particles at the outer section of the stream because the optical depth of the central part of the stream was too large for the motion of the particles to be tracked. Fig. 9 shows the cumulative percentage of the horizontal particle velocities of the 50-µm and 100-µm glass beads within 20.4 mm from the aperture in the vertical direction, and those for the alumina particles and silica sand within 19.4 mm and 17.9 mm from the aperture, respectively. The four types of particles had similar horizontal velocity distributions, typically about 1 cm s$^{-1}$.

According to Chokshi et al. (1993), the critical collision velocity between the monomers of radius $R$ is

$$v_{cr} \cong 3.86 \frac{\gamma^{\frac{5}{6}}}{[(1-\nu^2)/\varepsilon]^{-\frac{1}{3}} R^{\frac{5}{6}} \rho^{\frac{1}{2}}}, \quad (7)$$

where $\nu$ is the Poisson ratio, $\varepsilon$ is Young's modulus, and $\rho$ is density. On the other hand, we assume that agglomerates, not monomers, collide with each other in a granular stream, based on the fact that tiny agglomerates consisting of a small number of monomers are recognized in the images of Fig. 8. The critical collision velocity for equal-size agglomerates was shown to be

$$v_{cr} \cong 15 \sqrt{E_{\text{break}}/m}, \quad (8)$$
$$E_{\text{break}} \cong 23[\gamma^5 R^4 (1-\nu^2)^2/\varepsilon^2]^{1/3}, \quad (9)$$
$$v_{cr} \cong 35 \frac{\gamma^{\frac{5}{6}}}{[(1-\nu^2)/\varepsilon]^{-\frac{1}{3}} R^{\frac{5}{6}} \rho^{\frac{1}{2}}}, \quad (10)$$

where $E_{\text{break}}$ is the energy of breaking a single contact between two (identical) particles and *m* is the mass of the constituent monomer (Wada et al., 2013). Thus, the critical collision velocity between agglomerates is about nine times higher than the critical velocity between monomers. If we substitute the value of the effective surface energy, $\gamma_{\text{eff}}$, obtained in the previous section into $\gamma$ in Eq. (10), we obtain the critical velocities for the virtual agglomerates consisting of identical monomers with values of $R$ and $\gamma_{\text{eff}}$. Fig. 10 shows the calculated critical collision velocity for equal-size agglomerates versus the cumulative percentage. The critical collision velocity is a few cm s$^{-1}$. The maximum critical collision velocity between agglomerates exceeds the typical horizontal velocity of the particles in the stream (approximately 1 cm s$^{-1}$). Although we did not measure the typical horizontal velocity of the agglomerates, it would be similar to or lower than the individual particle horizontal velocity and therefore even lower than the maximum critical collision velocity between agglomerates. Consequently, agglomerates in a stream may grow via agglomerate-agglomerate collisions. The range of the critical velocity for agglomerates of each type of particle, based on Fig. 10, is shown in Table 3. The range of the critical collision velocity for agglomerates is indicated by the quadrangles in Fig. 9. We adopted the values of $\nu$ and $\varepsilon$ shown in Table 1. A relatively large number of the 50-μm glass beads had horizontal particle velocities lower than the critical collision velocity, indicating favorable conditions for agglomerate growth. On the other hand, most but not all of the 100-μm glass beads, alumina particles, and silica sand grains had higher horizontal velocities than the critical collision velocity. Agglomeration of these particles is expected to be more difficult.

**Observation of granular streams**
The state of granular streams at 0.9 m of vertical distance from the funnel aperture at each ambient pressure is summarized in Fig. 11. Because the transparent part of the vacuum chamber at Kobe University is 1.0 m high, the maximum drop height of the particles for taking images from outside of the chamber is approximately 0.9 m. In air, all particles except for the 100-μm glass beads formed millimeter-size agglomerates at 0.9 m. Although the 100-μm glass beads did not form agglomerates at 0.9 m from the funnel aperture, agglomeration was initiated at approximately 1.2 m from the aperture. Based on Figs. 9a and 9b, the 50-μm glass beads can form agglomerates more easily than the 100-μm glass beads. The observational results of the granular streams were in accordance with predictions based on the critical collisional velocity. Contrary to the prediction from the critical collision velocity, although the prediction was for spherical particles, alumina particles and silica sand grains were able to form agglomerates in air. A previous study showed that irregularly shaped particles have a higher probability of sticking and a higher capture velocity than spherical particles (Poppe et al., 2000). The reason was thought to be due to multiple opportunities of hitting and the resultant energy dissipation allowed by the irregular shape. Royer et al. (2009) also conducted laboratory experiments with irregularly shaped particles and found less agglomeration of irregular particles, a tendency opposite to that observed in the current study. The irregularly shaped particles used in Royer et al. (2009) were quite rounded, that is, not angular, while those used in this study were elongated and angular. This difference in irregularity would result in the different degrees of agglomeration. In fact, the irregularity of diamond particles of 1.3–1.9 μm in diameter (average diameter: 1.5 μm) used by Poppe et al. (2000) exhibiting higher sticking velocity than the spherical particles is similar to those used in our study, although the collision velocity in the current study was lower than that of Poppe et al. (2000) and the energy dissipation mechanism may have been different. Further study is required to reveal the formation process of the agglomerates in a granular stream.

Fig. 12 shows the ratio of the local minimum to the local maximum widths of granular streams, determined using ImageJ software versus the ambient pressure. When agglomerates form perfectly, the ratio becomes zero. As shown in Fig. 11, the formation of agglomerates becomes more difficult as the ambient pressure decreases. This is likely due to a decrease in the adhesive force, i.e., capillary forces, at lower pressures. Because the capillary force relates to the surface energy of water, which is 0.072 J/m$^2$ (Butt and Kappl, 2010), it contributes significantly to adhesive forces in air. However, there was no influence of the atmosphere in the previous study (Royer et al., 2009). The reason why the ambient gas pressure dependence in our study is different from that in the previous one remains unclear; however, it may be due to differences in the experimental conditions, such as the funnel aperture and humidity.

Fig. 13 shows the length ($\lambda$) and width ($w$) of agglomerates at 0.90 ± 0.01 m from the aperture in air. We used ImageJ software for the measurement. The length to width ratio was between approximately 1 and 2.5, which roughly agrees with the ratio (between 1 and 3) shown in the previous study (Royer et al., 2009); however, the absolute size of the agglomerates formed was larger. This does, however, agree with the correlation observed in the previous study (Royer et al., 2009): specifically, that a larger aperture leads to a larger agglomerate size. Agglomerates of irregular particles had a larger overall size than agglomerates of spherical particles.

**Application to planetary surfaces**
Crater rays have bright albedo features that extend from fresh craters, forming radially oriented filaments and diffuse patches (Melosh, 1989). Rays have been observed on airless bodies, i.e., the Moon, Mercury (Allen, 1977), and icy satellites (Wagner et al., 2011). Various processes, including deposition of ejecta from primary craters, have been proposed as the production mechanism of rays (Allen, 1977). Kadono et al. (2015) showed experimentally and numerically that clear mesh patterns form in impact-generated ejecta during flight. They also demonstrated in the laboratory that the characteristic angle of spaces between radial patterns was a few degrees, which is similar to that of the rays around two craters on the Moon. If crater rays are formed via ejecta from primary craters, the morphology of the rays can be discussed in terms of the agglomerating process of ejecta particles. The pattern of rays may be compared directly with the undulation of the granular streams shown in Fig. 11, although a dedicated study of the formation of discrete patterns in the ejecta curtain is required.

A study of agglomeration in a granular stream would also provide new insights into the origin of grooves on Phobos. The grooves, or striations, are often parallel depressions, and many have a pitted appearance (Veverka and Duxbury, 1977). Hypotheses regarding their origin include secondary impacts from the Stickney Crater, secondary impacts from primary impact events on Mars, impact crater chains from debris in Mars's orbit, tracks formed by boulders from the Stickney Crater, and fractures caused by impacts or tidal forces into which regolith has drained (Murchie et al., 2015). According to the hypothesis of secondary impacts from primary impact events on Mars, an ejecta melt jet derived from a large impact on Mars breaks up into droplets and causes a chain of impact craters on the surface of Phobos (Murray and Heggie, 2014). Even if melt droplets do not form, the ejecta particles may gather together, as seen in the granular stream of this study, impact Phobos with a beaded structure, and thus create pitted grooves. By comparing the aspect ratio of the pits of the grooves with that of the agglomerates in granular streams, this formation hypothesis may be evaluated.

Moreover, our study showed that both the adhesive force of particles and the particle shape greatly influence agglomerate formation. Tsuchiyama et al. (2011) showed that the shape distribution of the particles of asteroid (25143) Itokawa is not significantly different from that of laboratory experimental fragments, while the shapes of lunar regolith particles are more spherical than experimental impact fragments. Such information on the shape of ejecta particles may be obtained from ejecta curtains and crater rays originating from natural collisions or artificial collisions like the SCI experiment of Hayabusa2 (Arakawa et al., 2017; Ogawa et al., 2017) by applying the results of our study. The grain size of the surface of asteroid Ryugu, the target of Hayabusa2, is estimated to be 3-30 mm (Wada et al., 2018); however, there may be particles of smaller size, and even a depth-dependence of size and shape of regolith particles due to, for example, vibrational size sorting (Miyamoto et al., 2007) or thermal fatigue (Delbo et al., 2014) and alteration, caused by either solar radiation heating (Michel and Delbo, 2010) or space weathering (Le Corre et al., 2018). The macroscopic pattern of ejecta from the surface to subsurface may show a microscopic property of the constituent particles.

Finally, apart from its significance for planetary surfaces, understanding the agglomeration phenomenon in the granular stream may be useful for understanding particle shape effects of the agglomeration process of solid particles in protoplanetary disks and debris disks.

## Conclusions

We conducted laboratory experiments of granular streams using poly-dispersed glass beads with median diameters of 55 μm (50-μm glass beads) and 93 μm (100-μm glass beads), alumina particles, and silica sand grains to investigate the formation conditions of agglomerates in granular streams. The glass beads were spherical, whereas the alumina particles and silica sand grains had irregular, non-spherical shapes. We measured the adhesion of the particles using a centrifugal separation method. The 50-μm glass beads had a relatively smaller range of adhesion, which was closer to theoretical expectations than the other particles. The other particles showed discrepancies of orders of magnitude from the theoretical values of perfect spheres, as well as a large spread in the values, which was likely due to the microscopic roughness of the particle surfaces, in addition to the macroscopic irregularity in their shape. Based on the measured adhesion, the critical collision velocity for agglomerates of these particles was up to a few cm s$^{-1}$, while the typical horizontal velocity of the particles just below the aperture of the funnel was approximately 1 cm s$^{-1}$.

Agglomerate formation for the 50-μm glass beads occurred much more easily than for the 100-μm glass beads, which agrees with predictions based on a comparison between the horizontal velocity and critical collision velocity for agglomerates. However, irregularly shaped particles also formed agglomerates, although these showed less adhesion. It seems that more efficient energy dissipation occurs during the collisions among irregularly shaped particles. The sizes of the agglomerates were larger than in the previous study (Royer et al., 2009), likely due to the larger aperture of the funnel used in this study. The length and width ratios of the agglomerates were approximately between 1 and 2.5, which roughly agrees with the previous study (Royer et al., 2009), although it is notable that the agglomerates of irregular particles had a larger size than those of spherical particles.

The granular stream allows constituents to collide at velocities low enough for clusters or agglomerates to grow and therefore can be used to study the agglomeration of solid particles. Our study showed that both the adhesive force of particles and the particle shape greatly influence agglomerate formation. Therefore, information on the shape or adhesive force of ejecta particles may be obtained from ejecta curtains and crater rays originating from natural collisions or artificial collisions, like that of the SCI experiment of Hayabusa2. Further study of the agglomeration of solid particles in a granular stream would allow new insights into cratering ejecta patterns, rays on planetary bodies, and grooves on Phobos. Laboratory granular stream experiments may also be useful for examining the effects of the particle shape on other agglomeration processes, such as those in protoplanetary disks and debris disks.

## Abbreviations
ISAS: Institute of Space and Astronautical Science; JAXA: Japan Aerospace eXploration Agency; SEM: Scanning Electron Microscope, PIV: Particle Image Velocimetry

## Declarations
### Acknowledgement
We would like to thank K. Ogawa and A. Suzuki for their support with the laboratory experiments. We are also grateful to the Hypervelocity Impact Facility (former facility name: The Space Plasma Laboratory), ISAS, and JAXA for their support.

### Availability of data and materials
Please contact author for data requests.

### Competing interest
The authors declare that they have no competing interest.

### Funding
This research was supported by the Hosokawa Powder Technology Foundation, and grants-in-aid for scientific research from the Japanese Society for the Promotion of Science (No. 25400453 and No. 18K03723) of the Japanese Ministry of Education, Culture, Sports, Science, and Technology (MEXT).

### Author's contributions
YN conducted the measurements and experiments, analyzed the data, and constructed the manuscript. TO prepared the regolith analogues and helped measure the cohesive force of the particles. MK helped with the granular stream experiments. AMN proposed the topic, conceived the study, and helped in the interpretation of the results. KW provided the theoretical background of the study. SH supported the granular stream experiments conducted in the vacuum chamber in ISAS. All authors read and approved the final manuscript.

**Figure legends**

Figure 1. Particle size distributions. Cumulative volume fractions plotted against the particle diameter. (a) 50-μm glass beads, (b) 100-μm glass beads, (c) alumina particles, and (d) silica sand. Solid and dashed lines are the size distributions determined using a laser diffractometer and optical microscope images, respectively. In the latter, the diameter of the equivalent sphere of the projected area of the microscope image is used.

Figure 2. Scanning electron microscopy images of each particle: (a) 50-μm glass beads, (b) 100-μm glass beads, (c) alumina particles, and (d) silica sand grains. Scale bar is 500 μm.

Figure 3. (a) Schematic diagram of the experimental configuration used for the adhesive force measurements. (b) Image inside the centrifuge. There are four baskets (marked "B"), each of which can be tilted freely from vertical to horizontal. (c) Image inside the basket. A Petri dish of diameter 3.3 cm and height 1.5 cm was placed on a gypsum filler.

Figure 4. Optical microscope images of alumina particles (a) before and (b) after applying a centrifugal force at a rotation speed of 300 rpm. A particle in a circle in image (a) detached during the rotation and did not exist on the Petri dish in image (b).

Figure 5. Experimental setup of the granular stream experiments conducted at Kobe University: (a) overall view and (b) detailed view of the top part.

Figure 6. Measurement results of the adhesive force. The ratio of the measurement value of the adhesive force using a Petri dish to the theoretical reference value against the cumulative percentage. Solid, dashed, dotted, and dashed and dotted lines show the 50-μm glass beads, 100-μm glass beads, alumina particles, and silica sand, respectively. (a) The percentage reflects the number of detached particles/total number of detached particles ratio. (b) The percentage reflects the number of detached particles/total number of observed particles ratio.

Figure 7. Effect of surface roughness. The ratios of the measurement values of the adhesive force of the 100-μm glass beads obtained using a Petri dish and two optical glass plates having different surface roughnesses to the theoretical values plotted against the cumulative percentage. Solid line, dashed, and dotted lines correspond to the results of the Petri dish, smoother optical glass plate, and another optical glass plate, respectively. The numbers in parentheses show the roughness ($R_a$).

Figure 8. Granular streams at the funnel aperture in air: (a) 50-μm glass beads, (b) 100-μm glass beads, (c) 77-μm alumina particles, and (d) 73-μm silica sand.

Figure 9. Horizontal particle velocity distribution just below the funnel aperture plotted against cumulative percentage for (a) 50-μm glass beads, (b) 100-μm glass beads, (c) alumina particle, and (d) silica sand. Quadrangles show the ranges of the critical collision velocity for equal-size agglomerates corresponding to 10–90% of the total number of monomers.

Figure 10. Calculated critical collision velocity for equal-size agglomerates versus the cumulative percentage. This percentage is based on the number of detached particles/total number of detached particles ratio, as shown in Fig. 6a. Solid, dashed, dotted, and dashed and dotted lines show 50-μm glass beads, 100-μm glass beads, alumina particles, and silica sand, respectively.

Figure 11. State of granular streams at 0.9 m from the funnel aperture at each level of ambient pressure.

Figure 12. Ratio of the local minimum width to the local maximum width of granular stream versus ambient pressure.

Figure 13. Agglomerate length λ (mm) versus agglomerate width w (mm). Filled circle, open square, and triangle show 50-μm glass bead, silica sand, and alumina particles, respectively.

The solid line is the ratio $\lambda/w = 1$, and the dashed line is $\lambda/w = 3$.

Table 1 Physical characteristics of particles

| Particle type | Density (kg m$^{-3}$) | Surface energy (J m$^{-2}$) | Poisson's ratio, $\nu$ | Young's modulus, $\varepsilon$ (Pa) | Roughness, $Ra$ (μm) | Restitution coefficient |
|---|---|---|---|---|---|---|
| Glass bead 50 μm | $2.5 \times 10^3$ | 0.025[a] | 0.17[b] | $7.3 \times 10^{10}$[b] | $1.13 \pm 0.24$ | $0.957 \pm 0.013$ |
| Glass bead 100 μm | $2.5 \times 10^3$ | 0.025[a] | 0.17[b] | $7.3 \times 10^{10}$[b] | $3.82 \pm 0.45$ | $0.957 \pm 0.013$ |
| Silica sand 73 μm | $2.645 \times 10^3$ [c] | 0.025[a] | 0.17[b] | $7.3 \times 10^{10}$[b] | - | - |
| Alumina particle 77 μm | $4.0 \times 10^3$ | 0.041[d] | 0.24[e] | $3.85 \times 10^{11}$[e] | - | $0.975 \pm 0.006$ |

a Heim *et al.* (1999).

b Spinner (1962).

c The silica sand grains were used in a previous study (Omura et al., 2016), but we updated the density of the grain according to a measurement of true density made using a pycnometer (SEISHIN ENTERPRISE MAT-7000).

d Burnham *et al.* (1990).

e Wong and Bollampally (1999).

Table 2 Range of measured adhesive force between each particle and a Petri dish

| Particle type | Diameter range[a] (μm) | adhesive force[b] (μN) |
|---|---|---|
| Glass bead 50 μm | 44-66 | 0.16-5.6 |
| Glass bead 100 μm | 74-112 | 0.027-13 |
| Alumina particle 77 μm | 62-92 | 0.0062-13 |
| Silica sand 73 μm | 58-88 | 0.0069-8.0 |

a 80-120% of the median diameter (see Fig. 1).

b The corresponding range of values for the particles within the diameter range shown in the left column.

Table 3 Range of effective surface energy

| Particle type | $\gamma_{eff}$[a] (J m$^{-2}$) | $v_{cr}$[a] (cm s$^{-1}$) |
|---|---|---|
| Glass bead 50 μm | 0.00061-0.017 | 0.20-2.8 |
| Glass bead 100 μm | 0.000071-0.017 | 0.025-2.1 |
| Alumina particle 77 μm | 0.00012-0.011 | 0.022-0.98 |
| Silica sand 73 μm | 0.00034-0.010 | 0.099-1.7 |

aThe range of the $\gamma_{eff}$ and $v_{cr}$ values of the particles corresponding to 10–90% of the total number of particles.

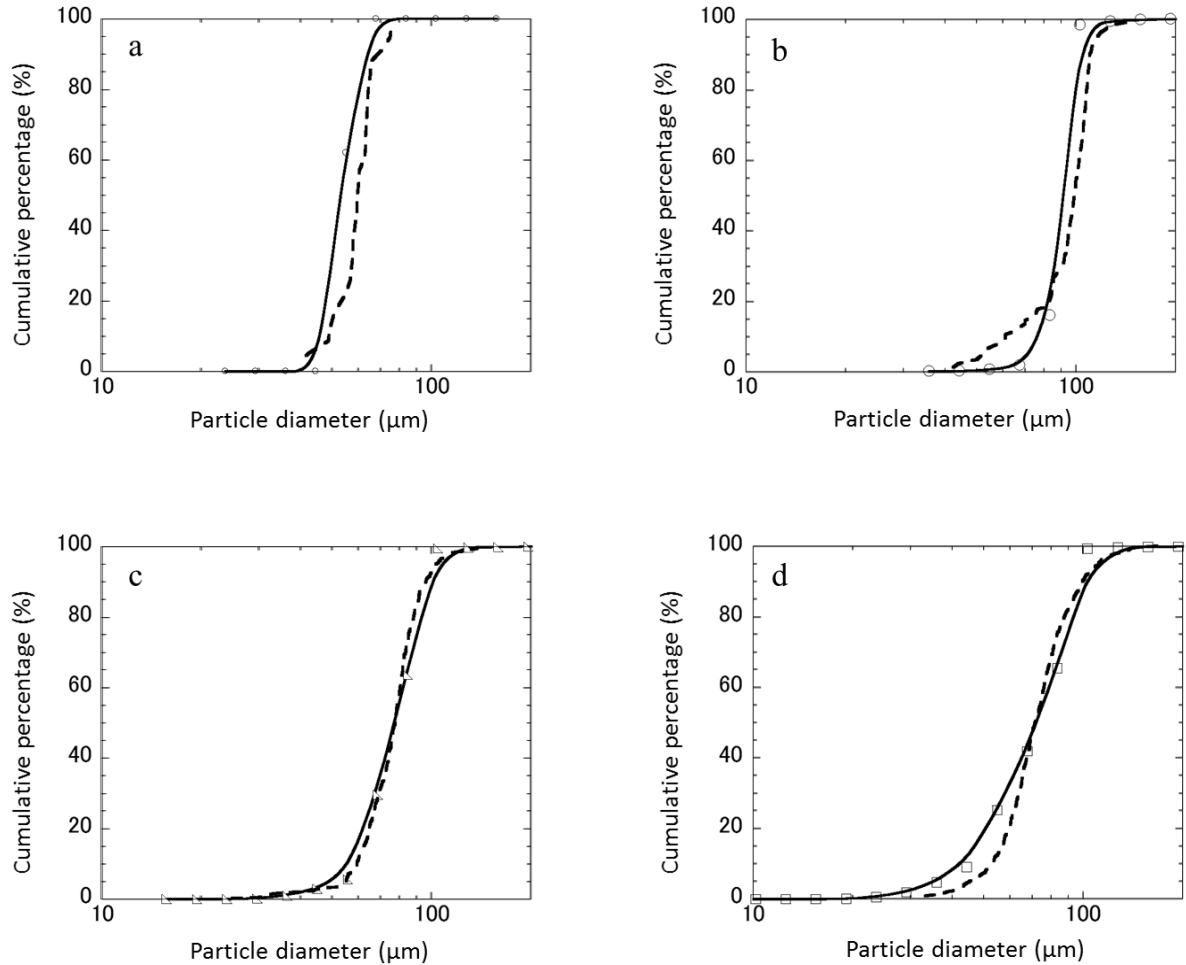

Figure 1. Particle size distributions. Cumulative volume fractions plotted against the particle diameter. (a) 50-μm glass beads, (b) 100-μm glass beads, (c) alumina particles, and (d) silica sand. Solid and dashed lines are the size distributions determined using a laser diffractometer and optical microscope images, respectively. In the latter, the diameter of the equivalent sphere of the projected area of the microscope image is used.

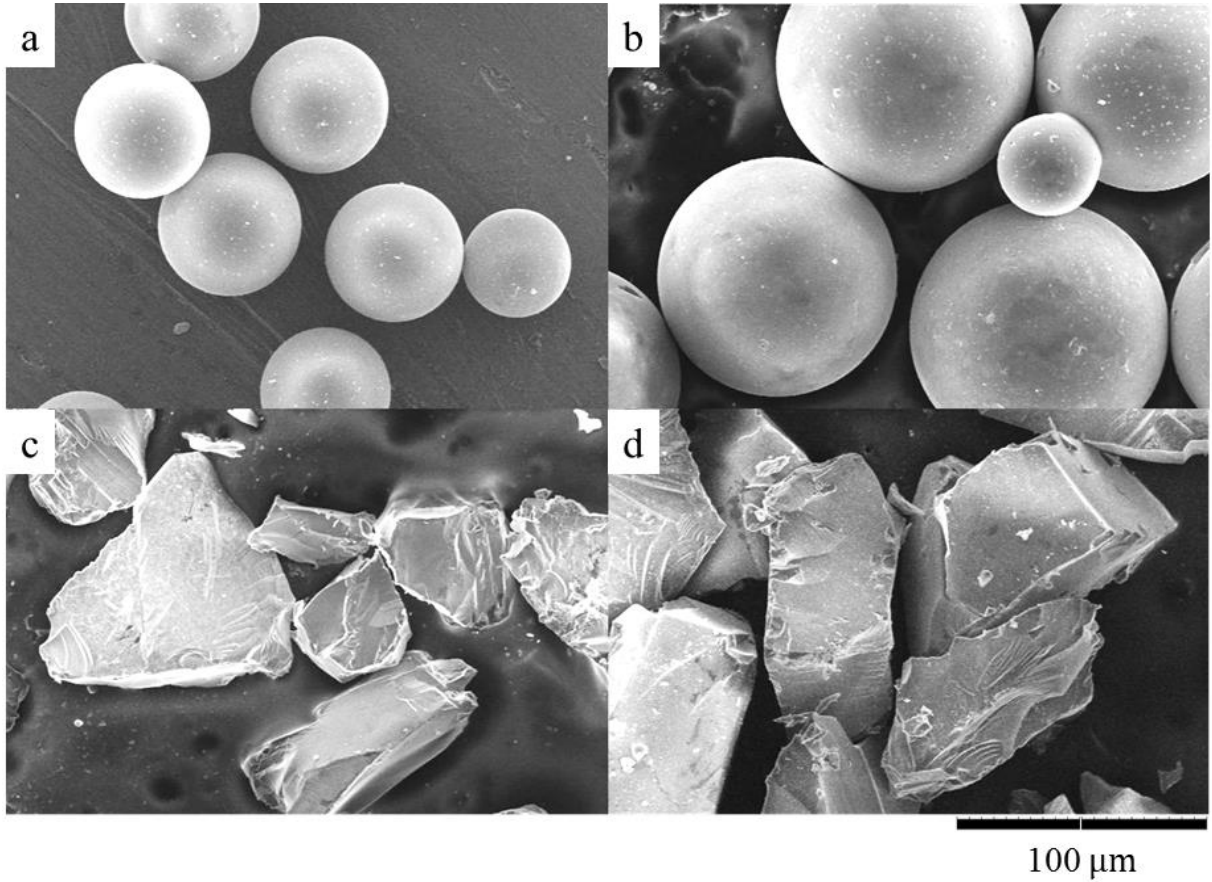

Figure 2. Scanning electron microscopy images of each particle: (a) 50-μm glass beads, (b) 100-μm glass beads, (c) alumina particles, and (d) silica sand grains. Scale bar is 500 μm.

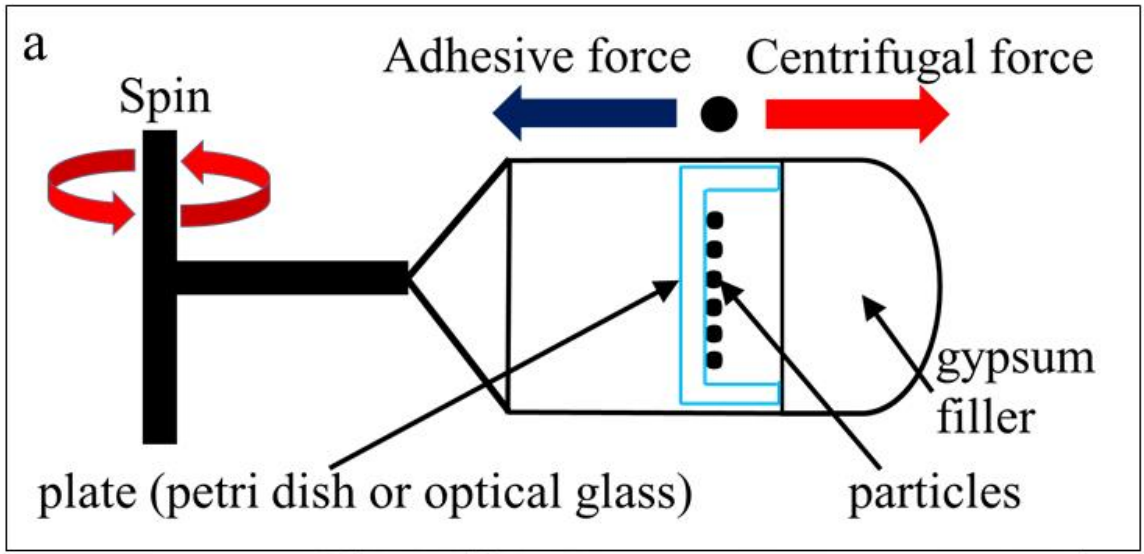
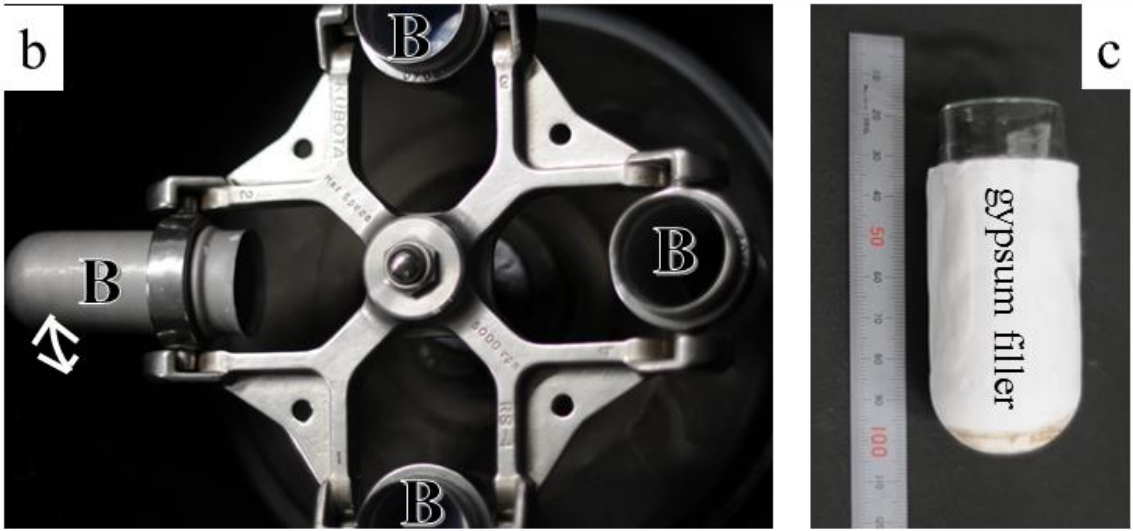

Figure 3. (a) Schematic diagram of the experimental configuration used for the adhesive force measurements. (b) Image inside the centrifuge. There are four baskets (marked "B"), each of which can be tilted freely from vertical to horizontal. (c) Image inside the basket. A Petri dish of diameter 3.3 cm and height 1.5 cm was placed on a gypsum filler.

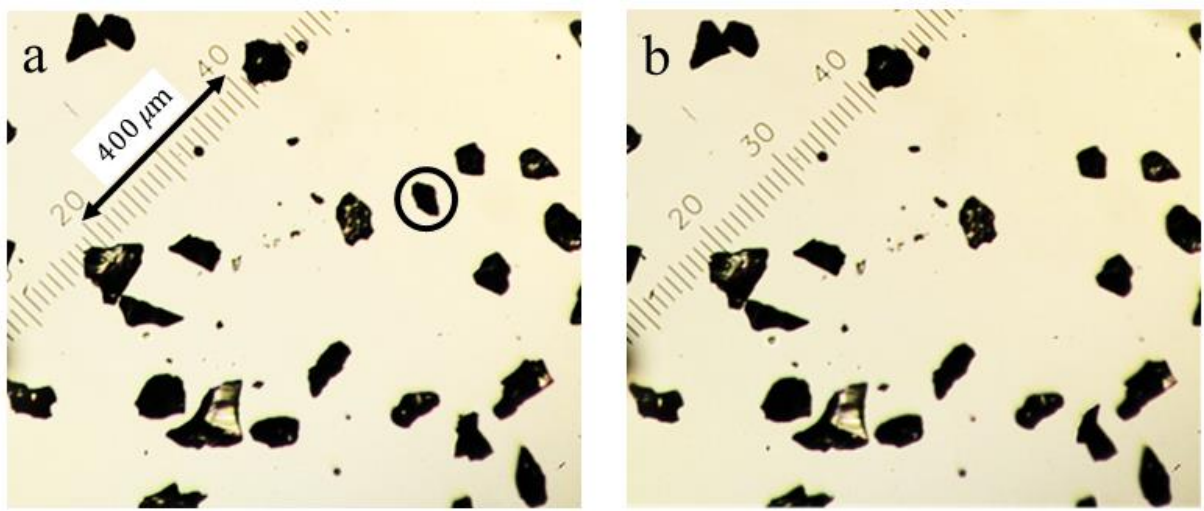

Figure 4. Optical microscope images of alumina particles (a) before and (b) after applying a centrifugal force at a rotation speed of 300 rpm. A particle in a circle in image (a) detached during the rotation and did not exist on the Petri dish in image (b).

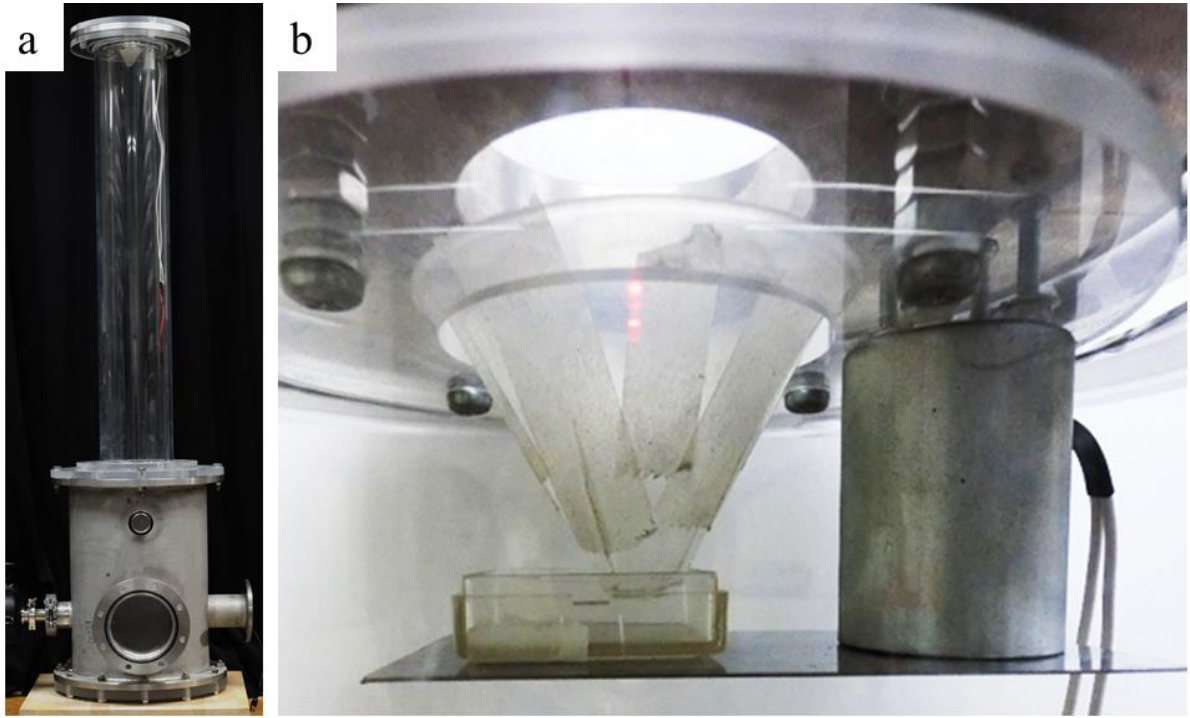

Figure 5. Experimental setup of the granular stream experiments conducted at Kobe University: (a) overall view and (b) detailed view of the top part.

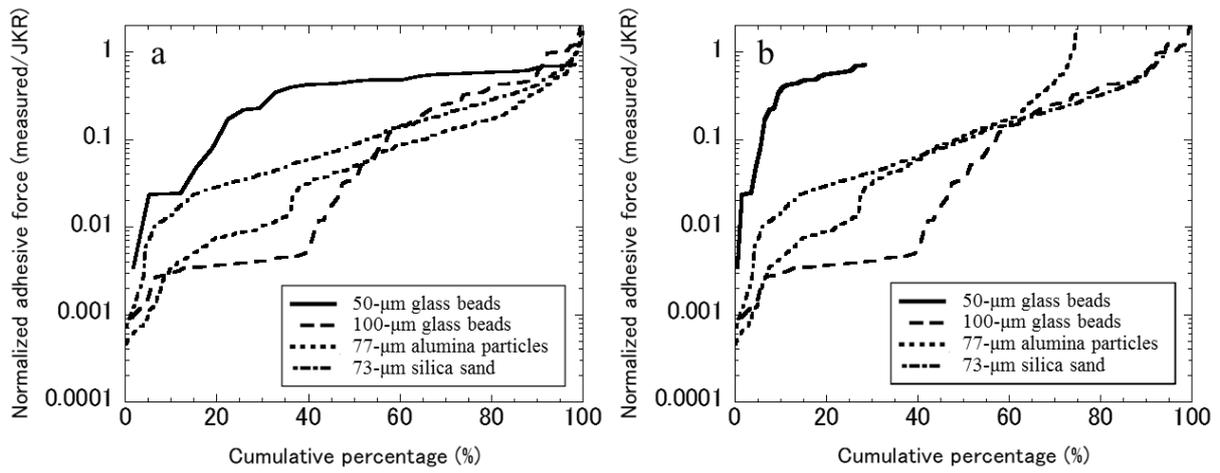

Figure 6. Measurement results of the adhesive force. The ratio of the measurement value of the adhesive force using a Petri dish to the theoretical reference value against the cumulative percentage. Solid, dashed, dotted, and dashed and dotted lines show the 50-μm glass beads, 100-μm glass beads, alumina particles, and silica sand, respectively. (a) The percentage reflects the number of detached particles/total number of detached particles ratio. (b) The percentage reflects the number of detached particles/total number of observed particles ratio.

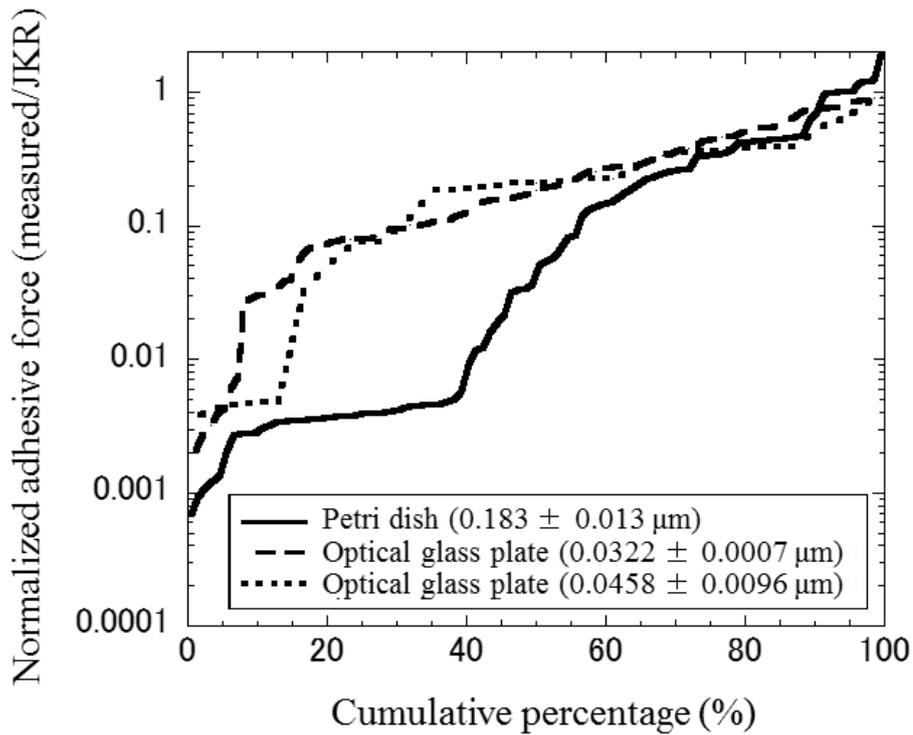

Figure 7. Effect of surface roughness. The ratios of the measurement values of the adhesive force of the 100-μm glass beads obtained using a Petri dish and two optical glass plates having different surface roughnesses to the theoretical values plotted against the cumulative percentage. Solid line, dashed, and dotted lines correspond to the results of the Petri dish, smoother optical glass plate, and another optical glass plate, respectively. The numbers in parentheses show the roughness ($R_a$).

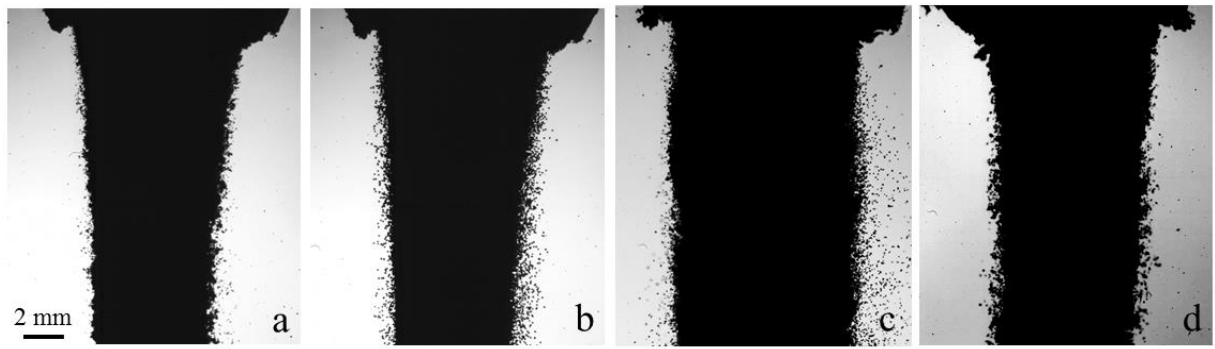

Figure 8. Granular streams at the funnel aperture in air: (a) 50-μm glass beads, (b) 100-μm glass beads, (c) 77-μm alumina particles, and (d) 73-μm silica sand.

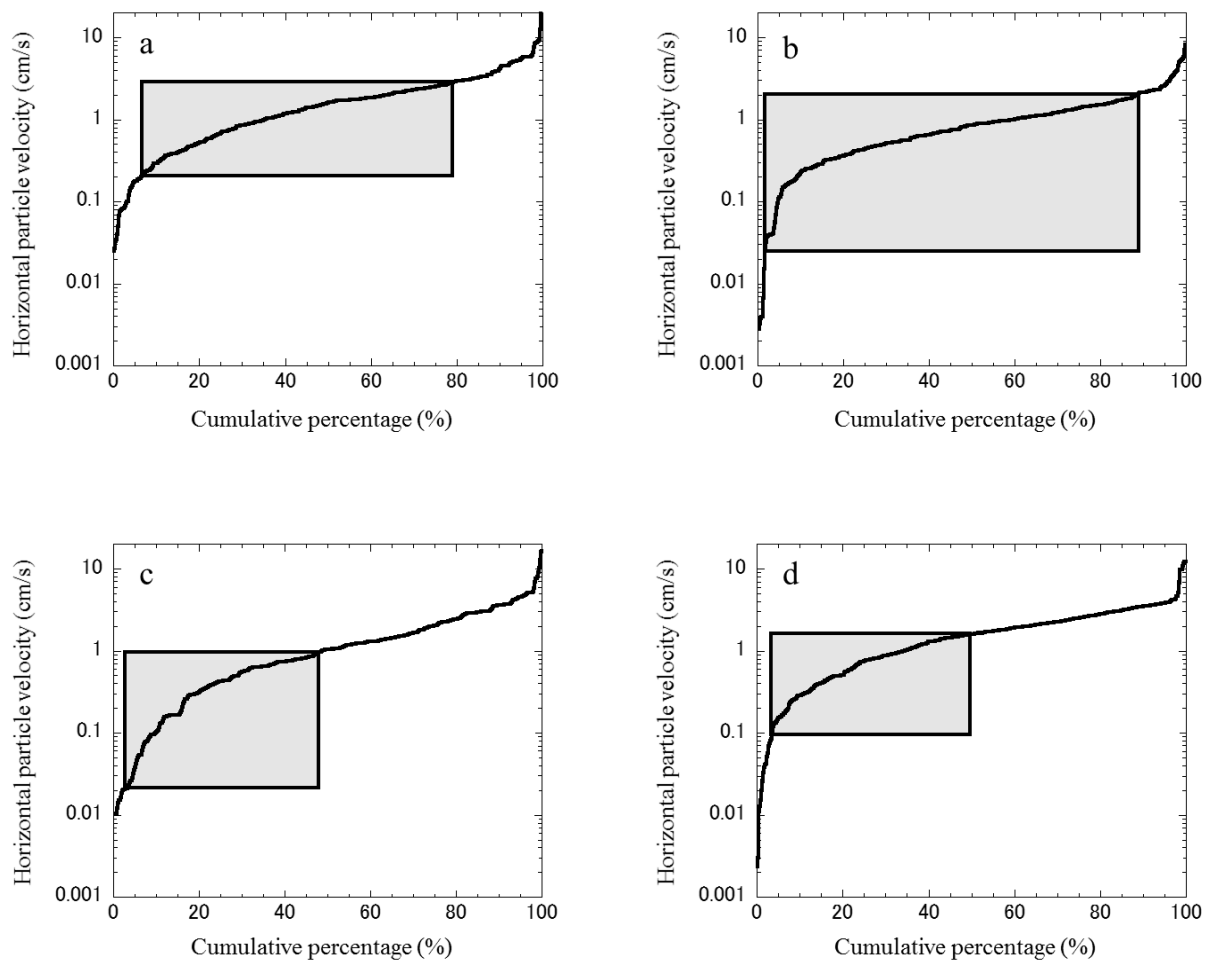

Figure 9. Horizontal particle velocity distribution just below the funnel aperture plotted against cumulative percentage for (a) 50-μm glass beads, (b) 100-μm glass beads, (c) alumina particle, and (d) silica sand. Quadrangles show the ranges of the critical collision velocity for equal-size agglomerates corresponding to 10–90% of the total number of monomers.

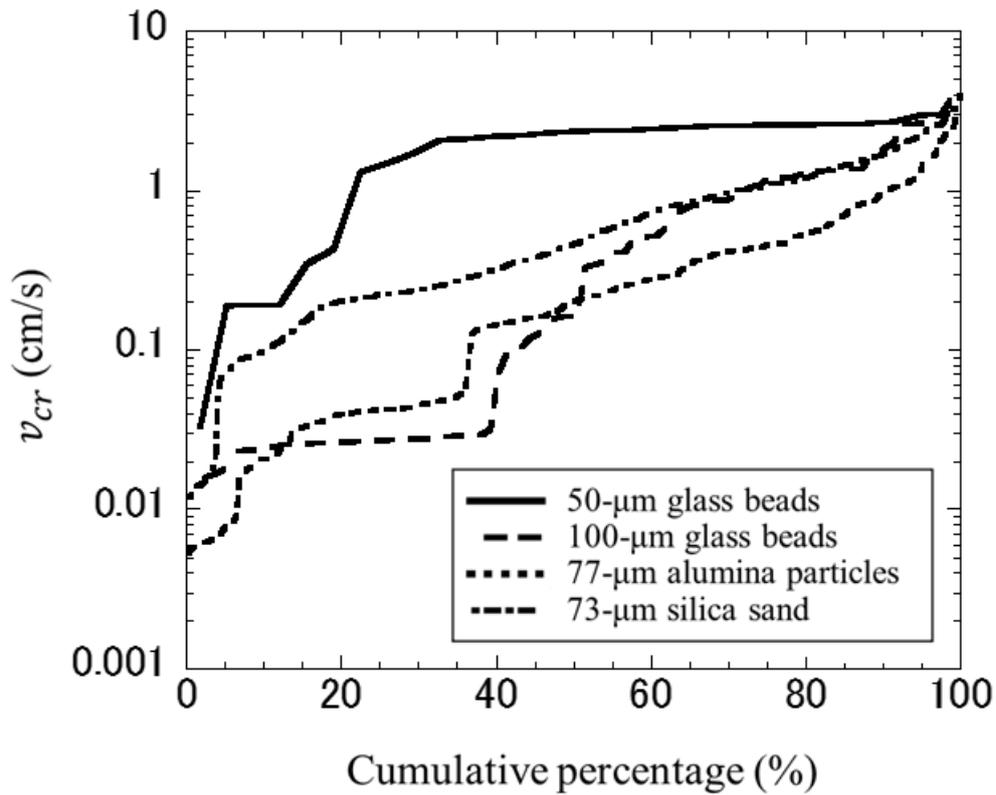

Figure 10. Calculated critical collision velocity for equal-size agglomerates versus the cumulative percentage. This percentage is based on the number of detached particles/total number of detached particles ratio, as shown in Fig. 6a. Solid, dashed, dotted, and dashed and dotted lines show 50-μm glass beads, 100-μm glass beads, alumina particles, and silica sand, respectively.

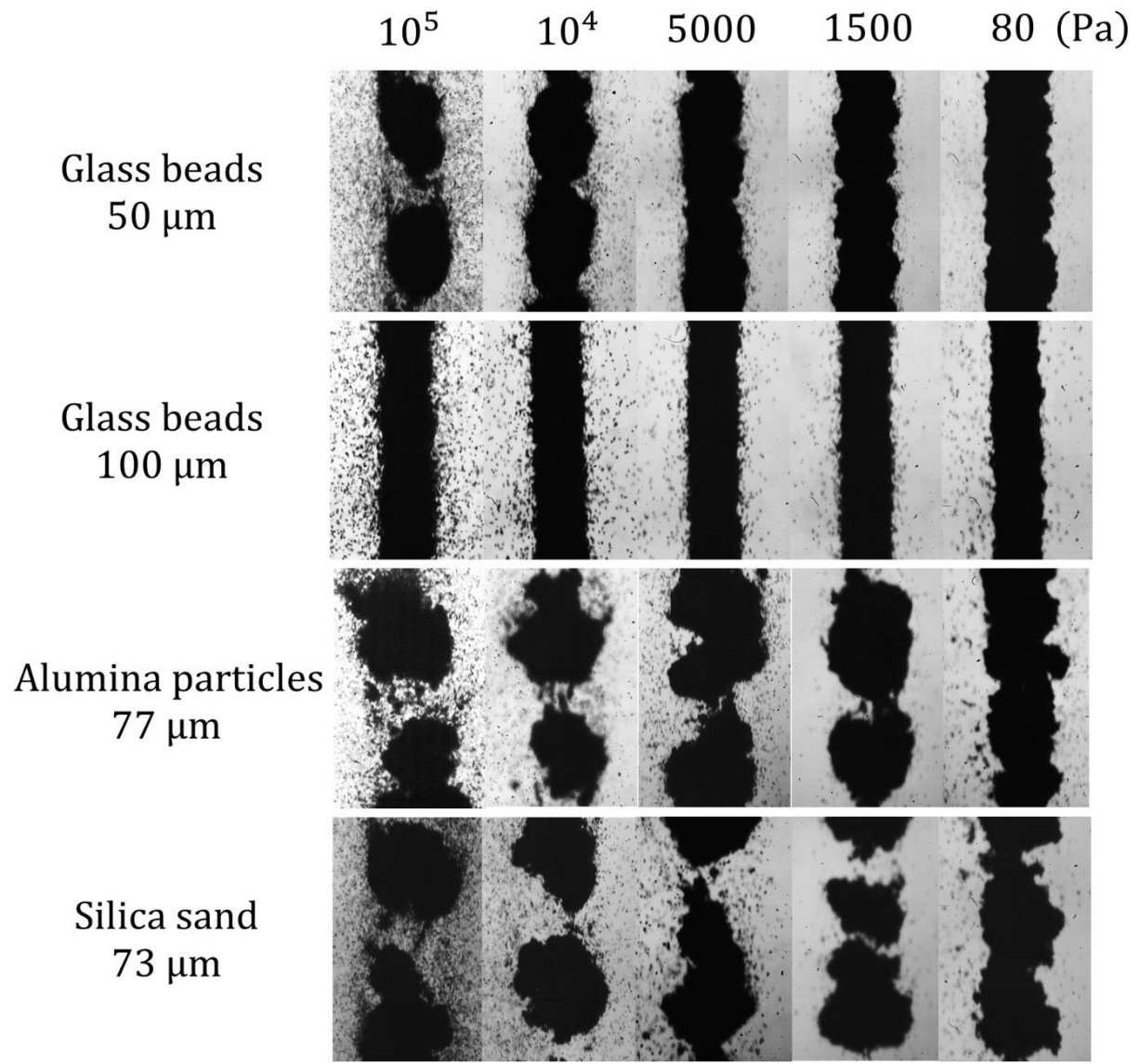

Figure 11. State of granular streams at 0.9 m from the funnel aperture at each level of ambient pressure.

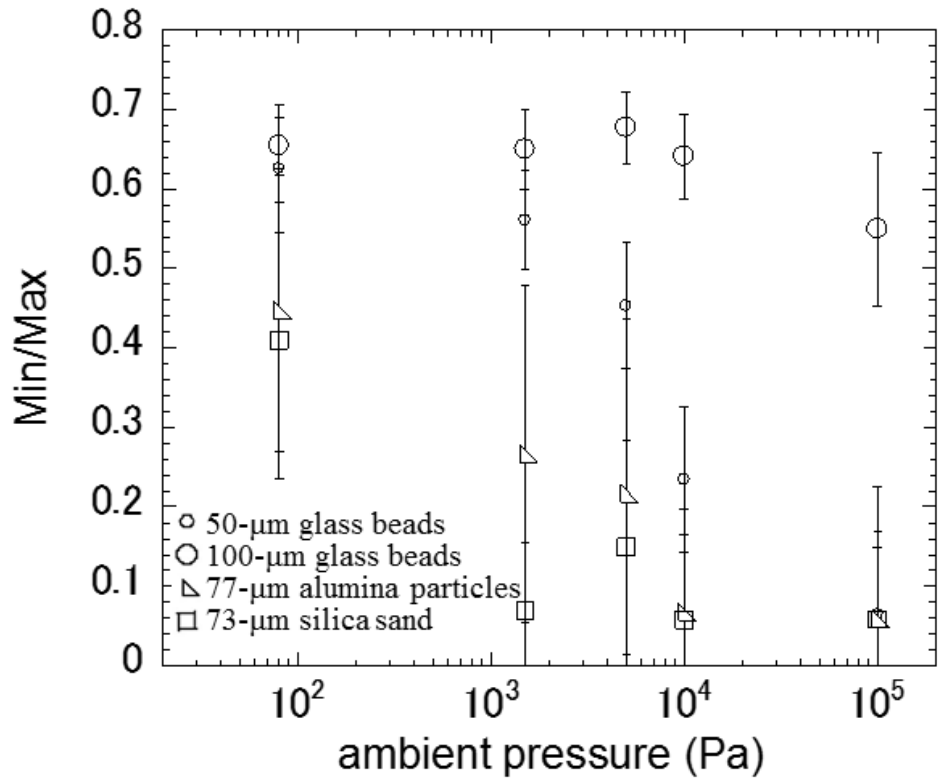

Figure 12. Ratio of the local minimum width to the local maximum width of granular stream versus ambient pressure.

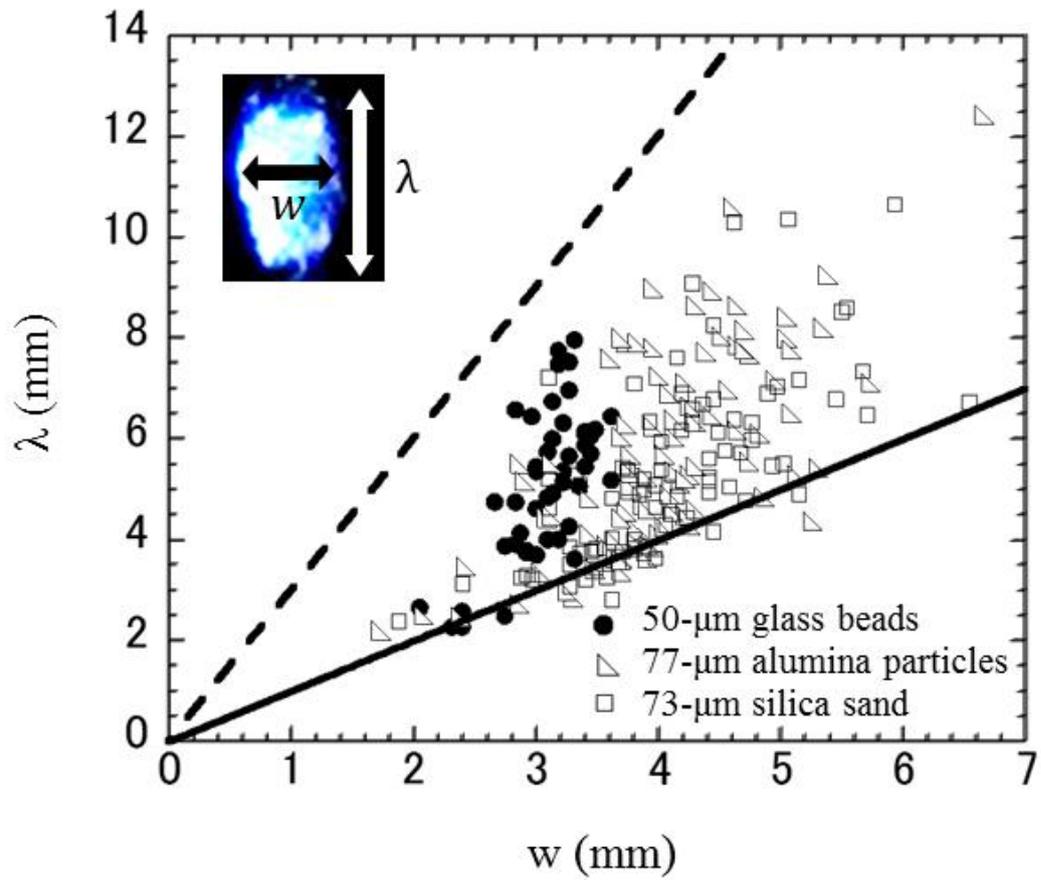

Figure 13. Agglomerate length λ (mm) versus agglomerate width w (mm). Filled circle, open square, and triangle show 50-μm glass bead, silica sand, and alumina particles, respectively.